\documentclass[a4paper,fleqn,usenatbib]{mnras}

\usepackage{newtxtext,newtxmath}

\usepackage[T1]{fontenc}
\usepackage{ae,aecompl}

\usepackage{graphicx,subfigure}	
\usepackage{amsmath}	
\usepackage{textgreek}  
\usepackage{placeins} 
\usepackage{wasysym} 
\usepackage{soul,xcolor} 
\usepackage[colorinlistoftodos]{todonotes}

\title{Low-redshift quasars in the SDSS Stripe 82 - III MOS observations}

\author[D. Bettoni]{
D. Bettoni,$^{1}$\thanks{E-mail: \href{mailto:daniela.bettoni@inaf.it}{daniela.bettoni@inaf.it}}
R. Falomo$^{1}$
S. Paiano$^{2}$
J. K. Kotilainen$^{3,4}$
and M. B. Stone$^{3,4}$ 
\\
$^{1}$INAF - Osservatorio Astronomico di Padova, Vicolo dell'Osservatorio 5, I-35122 Padova (PD), Italy\\
$^{2}$INAF - IASF Palermo via Ugo La Malfa 153,I-90146, Palermo Italy \\
$^{3}$Finnish Centre for Astronomy with ESO (FINCA), University of Turku, 20014 Turku, Finland \\
$^{4}$Department of Physics and Astronomy, University of Turku, 20014 Turku, Finland
}

\date{Accepted XXX. Received YYY; in original form ZZZ}

\pubyear{2022}

\begin{document}
\label{firstpage}
\pagerange{\pageref{firstpage}--\pageref{lastpage}}
\maketitle

\setstcolor{red}

\begin{abstract}
\null \\
We present multi object optical spectroscopy of the galaxies in the environment of 12 low-redshift (z~$<$~0.5) quasars and of 11 inactive massive galaxies chosen to match the properties of the quasar host galaxies to probe physical association and possible events of recent star formation.
The quasars are selected from a sample of QSOs in the SDSS Stripe82 region for which both the host galaxy and the large scale environments were previously investigated. 
The new observations complement those reported in our previous works on close companion galaxies of nearby quasars. For the whole dataset  we find that for about half (19 out of 44 ) of the observed QSOs there is 
at least one associated companion galaxy. In addition to the new spectroscopic observations, we add data from the SDSS database for the full sample of objects.
We find that the incidence of companion galaxies in the fields of QSO (17\%) is not significantly different from that of inactive galaxies  (19\%) similar to quasar hosts in redshift and mass.
Nevertheless, the companions of quasars exhibit more frequently emission lines than those of inactive galaxies, suggesting a moderate link between the nuclear activity and recent star formation in their environments.

\end{abstract}

\begin{keywords}
galaxies: active -- galaxies: evolution -- galaxies: nuclei -- ({\it galaxies:}) quasars: general.
\end{keywords}


\defcitealias{Bettoni_2017}{I}
\defcitealias{Falomo_2014}{F14}

\section{Introduction}

The nuclear activity in quasars is assumed to occur due to a major merger of two gas-rich galaxies that feed the central engine and enable the growth of a stellar spheroid. However, details on what triggers the gas fuelling and how nuclear activity affects the subsequent evolution of the host galaxies remain not fully understood. The correlations observed between the black hole (BH) mass ($M_{BH}$) and the properties of the associated galaxy bulge as the $M_{BH}$-$\sigma_{stars}$ \citep{Ferrarese_2000} or $M_{BH}$-$M_{bulge}^{*}$ \citep{Kormendy_2013} seems to point to a co-evolution of BH and spheroids but it did not yet clarify the role played by the environment, which is a fundamental component in investigating issues of quasar activity and its role in the evolution of galaxies. Minor and major mergers may have a key role for triggering and fuelling the nuclear activity. The global properties of the galaxy environment are probably the main driver of the AGN activity \cite[i.e.][]{Kauffmann_2000,DiMatteo_2005}.
  
However, the role of the environment on nuclear activity is still a puzzling subject; in fact, it is relevant to constrain processes that trigger and control activity and also to investigate its relationship to star formation. For example, the relationship of galaxy environment with star formation and/or morphology may not be universal \citep[e.g.][]{Wijesinghe_12}.

Looking at the environment at Mpc scales, comparing that of quasars to those of galaxies has given conflicting results partially due to limited samples and lack of homogeneous datasets. Early studies on the galaxy environments suggest that quasars are more strongly clustered than galaxies \citep[e.g.][]{Shanks_88}, while later studies based on surveys, such as the Two Degree Field (2dF) survey and the Sloan Digital Sky Survey (SDSS), found galaxy densities around quasars and inactive galaxies to be comparable to each other \citep[e.g.][]{Smith_2000,Wake_2004}. Using the SDSS archives \citet{Serber_06} and \citet{Strand_08} have taken advantage of the large datasets provided by the survey to study quasars at z $<$ 0.4. Both studies found that quasars are located in higher local over-density regions than typical L* galaxies, and that density enhancement is strongest
within 100 kpc from the quasar. \citet{Serber_06} also found that high luminosity quasars
have denser small-scale environments than QSOs at lower luminosity. This result was pointed out also by \citep{Shen_09} who find that at z$<$2.5 the 10\% of the most optically luminous quasars are more strongly clustered. More recently, \citet{Stott_20} found that UV bright QSO in the same range of redshift are found to be in overdense regions.
In spite of these studies our knowledge of the physical association of the galaxies around  low-z QSO and their dynamical properties  still remain poor. However, \citet{Karhunen_2014} finds that the number density of galaxies in a projected volume of 1 Mpc radius for a sample of QSO ($z<0.5$) and a comparison sample of passive galaxies matched in luminosity with the host galaxy of the QSO is very similar. More recently \citet{Wethers_2022} analyzing data from the Galaxy and Mass Assembly survey (GAMA) found again that QSO activity weakly depends on the galaxy environment both on large (Mpc) and small ($<$ 100 kpc) scales.

In the past years, a series of papers \citep{Falomo_2014,Karhunen_2014,Bettoni_2015} studied the host galaxy properties, environment and star formation (SF) for a large ($\sim$ 400) sample of low-redshift (z $<$ 0.5) quasars in the SDSS Stripe 82 area. In particular, with regard to the environment, \citet{Karhunen_2014} found that quasars are on average found associated with small group of galaxies. The over-densities of galaxies are mainly observed in the closest ($< $200 kpc) region around the source and vanish beyond a distance of 1 Mpc. The crucial test for this result is the spectroscopic study of the galaxies in the closest volume . We have recently completed a long-slit spectroscopic study \citep{Bettoni_2017,Stone_2021} for 34 QSO in which we found that $\sim$44\% have at least one associated galaxy. We found that many of the associated companions and some host galaxies exhibit episodes of (recent) star formation possibly induced by past interactions \citep{Stone_2021}. The star formation rate (SFR) of the companion galaxies is, however, modest, and the role of the quasar remains uncertain. The long-slit spectroscopy, of course, get an incomplete view of the environment being limited to the very close companions  and is very telescope time expensive. Here we extend the sample in order to gain statistical significance and thanks to the MOS observations to gain a more complete view of the close environment. In addition, in this paper we add a further crucial step of obtaining similar data for inactive galaxies well matched in redshift and galaxy luminosity to the quasar hosts.

This paper is organized as follows. 
The data sample is presented in Section 2 and the analysis in Section 3. The results of this study and an overview including findING from \cite{Bettoni_2017}, hereafter Paper \citetalias{Bettoni_2017}, and \cite{Stone_2021} hereafter Paper II, are presented in Section 4. Finally, in Section 5, we summarize our main conclusions.
The results are obtained in the framework of the concordance cosmology, using $ H_{0}$ = 70 km s$^{-1}$ Mpc$^{-1}$, $ \Omega_{m} $= 0.3, $ \Omega_{\lambda}$ = 0.7. 

\section{The sample}

\begin{figure}
    \includegraphics[width = 8cm]{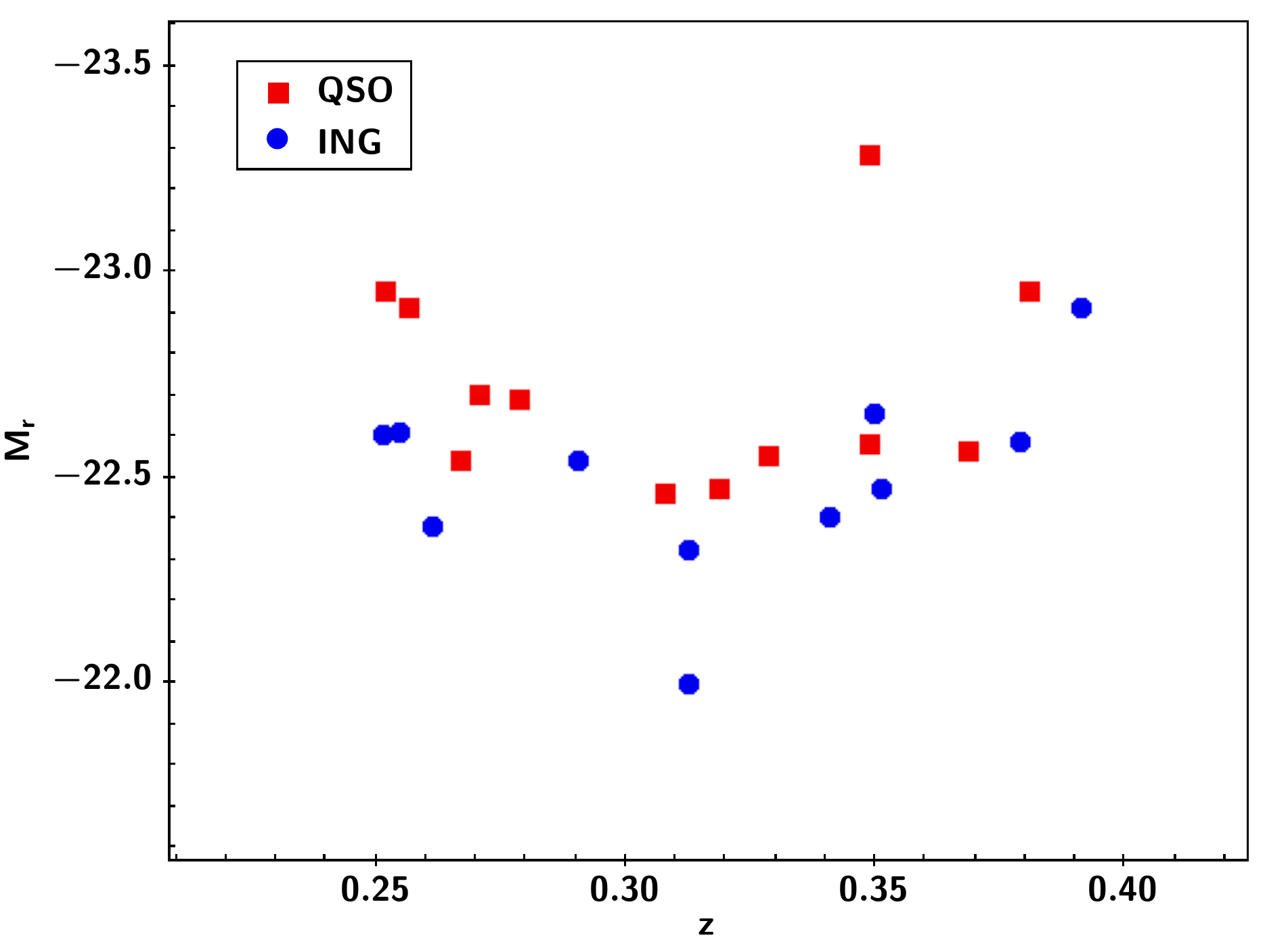}
\caption{Scatter plot of the absolute $r$ magnitude as a function of the redshift of the QSO host galaxies and of the ING. }
\label{fig:scatter}
\end{figure}

We obtained spectra of the objects in the fields of 12 QSO and 11 inactive galaxies (hereinafter ING). 
The QSO targets were selected from the \cite{Falomo_2014} parent sample of 416 low-redshift QSO located in the Stripe 82 \citep{Annis_2014} region of the sky for which complete information was provided concerning the characterization of the host galaxy and their galaxy environments. The QSO targets for this study were selected on the basis of good observability and feasibility from the adopted instrumentation. The selected fields are also chosen among objects well resolved (in \citet{Falomo_2014} objects defined as {\it resolved}) in order to study the possible connection between host and associated companions and events of recent SF. For each field, the number of possible observed companions was limited to 8-11, as to not overlap slitlets on the MOS plate, see also \S3.
For the comparison sample of ING, we selected the targets from the study of 580 passive galaxies by \cite{Karhunen_2014}. This  control sample of galaxies has the same  redshift  distribution of the QSO sample. Moreover we also matched the galaxy absolute  magnitude  to that of the host galaxy of the  resolved  QSO (see  \citet{Falomo_2014} for details). 
The average redshift is z=$<0.31>\pm 0.04$ and z=$<0.32>\pm 0.05$ for QSOs and ING, respectively and the average absolute magnitude M$_r = <-22.72>\pm 0.24$ and M$_r =<-22.49>\pm 0.22$ (see also Figure \ref{fig:scatter}).

The two samples are also comparable in terms of large scale environment properties \citep{Karhunen_2014}. The galaxy number density within 0.25 Mpc is 7.1 $\pm$ 2 for QSOs and 7.0 $\pm$ 3 for ING. 
The lists of the observed quasar and ING fields are given in Tables \ref{tab:obs_QSO} and \ref{tab:obs_GAL}, respectively. 
 
\begin{table}
    \centering
    \caption{Observed QSO fields}
    \label{tab:obs_QSO}
    \begin{tabular}{lcccccc}
        \hline
  \multicolumn{1}{c}{Nr} &    \multicolumn{1}{c}{RA} &    \multicolumn{1}{c}{Dec} &
    \multicolumn{1}{c}{z} &       \multicolumn{1}{c|}{$M_r$} &   \multicolumn{1}{c|}{$M_r$} &
     \multicolumn{1}{c}{$N_{slit}$} \\
  & 2000 & 2000 & & (nuc) & (host)& \\
  \hline
16 & 312.4859 & -0.20048 & 0.3693 & -24.5 & -22.56 & 4\\
  62 & 329.4341 & 0.88434 & 0.2674 & -22.23 & -22.54 & 5  \\
  68 & 329.95421 & 0.16799 & 0.2713 & -22.93 & -22.7  & 5 \\
 127 & 348.21203 & 0.28862 & 0.257 & -22.41 & -22.91 & 7 \\
 147 & 355.42293 & -0.63519 & 0.319& -23.85 & -22.47  & 4 \\
 154 & 357.38655 & -0.61273 & 0.279& -23.63 & -22.69  & 4 \\
 192  & 7.13215 & -0.07036 & 0.2519& -23.03 & -22.95  & 16 \\
 195  & 8.63226 & -0.2202 & 0.3811 & -23.51 & -22.95 & 6 \\
 204  & 10.5561 & 0.98825 & 0.3289 & -22.95 & -22.55 & 4 \\
 205  & 10.83224 & 0.85429 & 0.3083& -23.28 & -22.46  & 10 \\
 239  & 20.21226 & -0.30916 & 0.49 & -22.9 & -23.28 & 15 \\
 270  & 28.16225 & 0.15987 & 0.3492& -23.2 & -22.58  & 8 \\
        \hline
    \end{tabular}
     \begin{list}{}{}
    \item[]Column (1) catalog identification number from \cite{Falomo_2014}, column (2) object RA, column (3) object Dec, column (4) redshift, columns (5) and (6) k-corrected absolute r magnitude of the QSO and of the host galaxy and column (7) Number of used slits in the field.
    \end{list} 
    \end{table}
    
    \begin{table}
    \centering
    \caption{Observed ING fields}
        \label{tab:obs_GAL}
    \begin{tabular}{lccccc}
        \hline
  \multicolumn{1}{c}{Nr} &   \multicolumn{1}{c}{RA} &   \multicolumn{1}{c}{Dec} &
    \multicolumn{1}{c}{z} &       \multicolumn{1}{c|}{$M_r$} &    \multicolumn{1}{c}{$N_{slit}$} \\
  & 2000 & 2000 & & \\
  \hline
 32  &  6.13655 &  0.136926 & 0.3793 & -22.58 & 4 \\
 35  &  7.256614 & -0.47575 & 0.2906 & -22.54 & 5 \\
 45  &  10.18925 & 0.14944 & 0.313 & -22.32 & 2 \\
 55  &  12.601671 & -0.381672 & 0.2517 & 22.60 & 5 \\
 76  &  16.32155 &  0.898009 & 0.2616 & -22.38 &  9 \\
 77  &  16.4306 &  0.8878843 & 0.3516 & -22.47 & 4 \\
 80  &  16.54892 & -0.3190616 & 0.3516 & -21.40 & 4 \\
 81  &  16.55616 & 0.9371725 & 0.3913 & -22.91 & 4 \\
 120 & 16.55616 &  0.9371725 & 0.3913 & -22.00 & 2 \\
 514 & 337.48396 & 0.070315 & 0.2549 & -22.61 & 5 \\
 548 & 350.21705 & -0.301583 & 0.3499 & -22.66 & 9 \\
        \hline
    \end{tabular}
      \begin{list}{}{}
    \item[]Column (1) catalog identification number from \cite{Karhunen_2014}, column (2) object RA columns. (3) object Dec, column (4) redshift, column (5) k-corrected absolute r magnitude and (6) Number of used slits in the field.
    \end{list} 
    \end{table}

\begin{figure*}
    \includegraphics[width = 18cm]{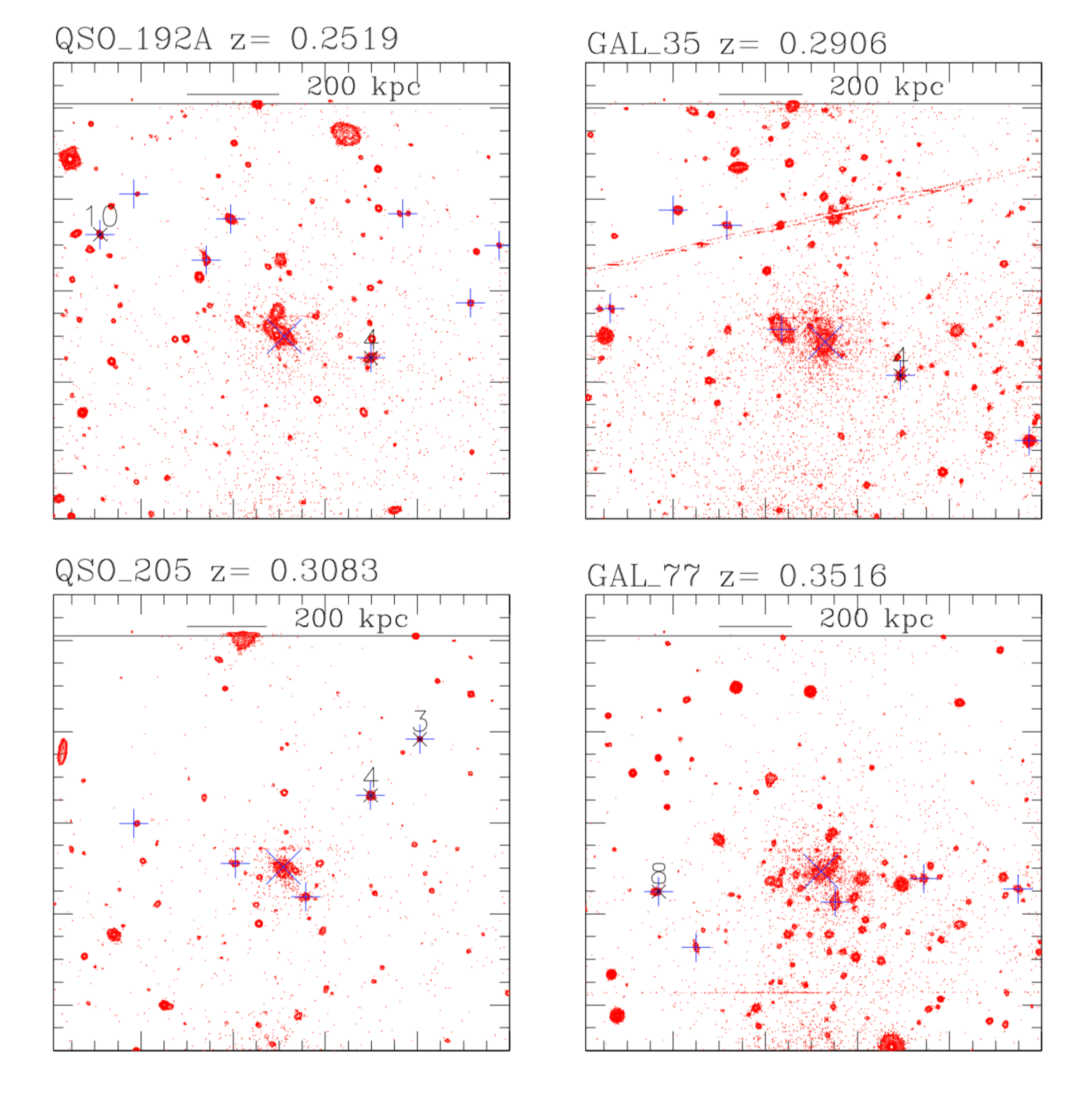}
\caption{$r$ band contour plot of observed QSO and ING fields. 
The numbers of QSO and galaxies refer to objects in \protect\citet{Falomo_2014} and \citet{Karhunen_2014}. The large blue X in the center indicates the target. Companion galaxies are labelled by a star and the slit number. The other slits observed are marked with blue crosses. Only the slits for the spectra with enough signal-to-noise are plotted. The top bar shows a scale of 200 kpc at the redshift of the target. North is up and East to the right.
}
\label{fig:ima}
\end{figure*}   

\begin{figure*}
      \includegraphics[width = 18cm]{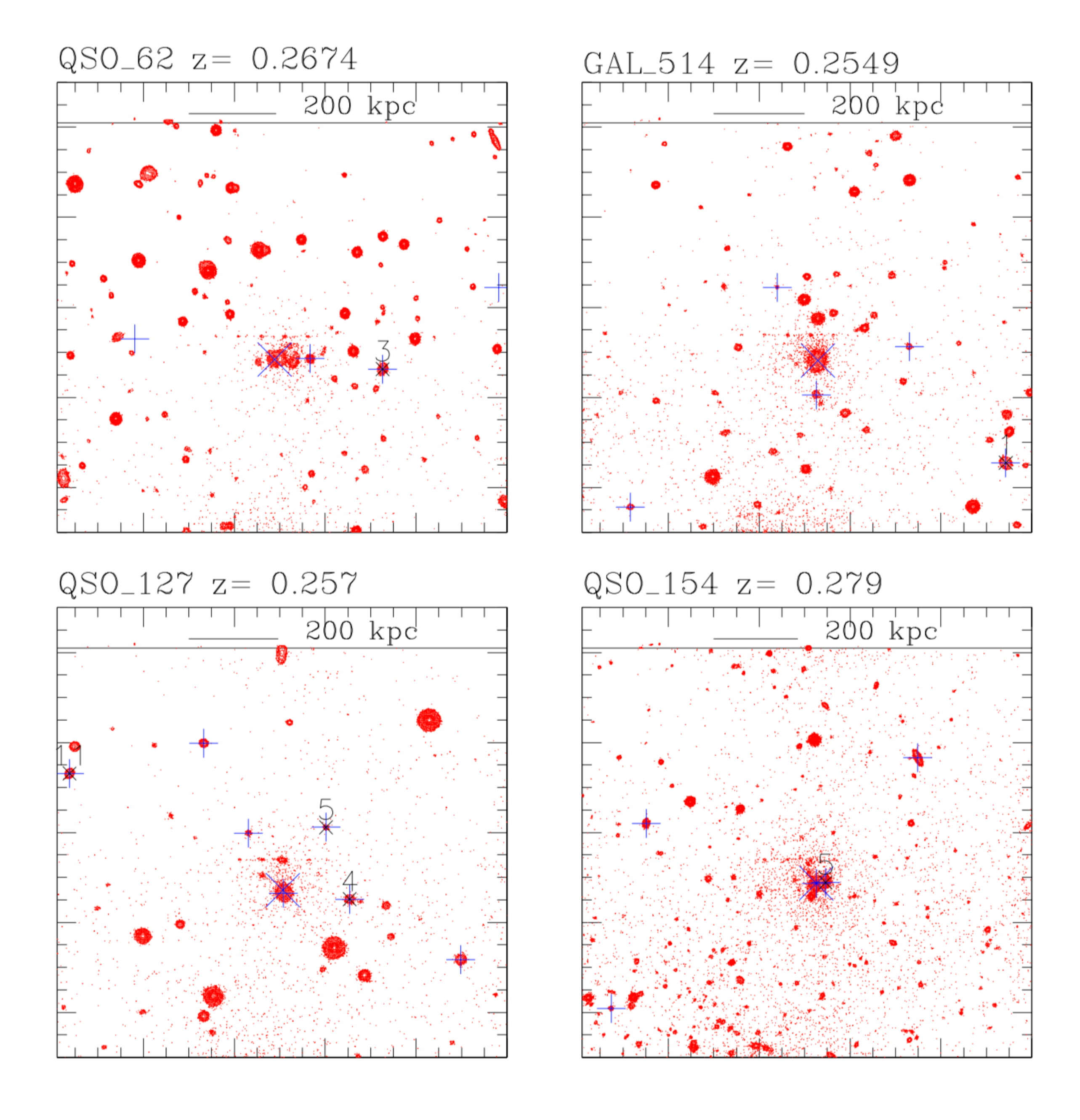}
\caption{$r$ band contour plot of observed quasar and ING fields. 
The numbers of QSO and galaxies refer to objects in \protect\citet{Falomo_2014} and \citet{Karhunen_2014}. The large blue X in the center indicates the target. Companion galaxies are labelled by a star and the slit number. The other slits observed are marked with blue crosses. Only the slits for the spectra with enough signal-to-noise are plotted. The top bar shows a scale of 200 kpc at the redshift of the target. North is up and East to the right. }
\label{fig:ima1}
\end{figure*}   

\begin{figure*}
      \includegraphics[width = 18cm]{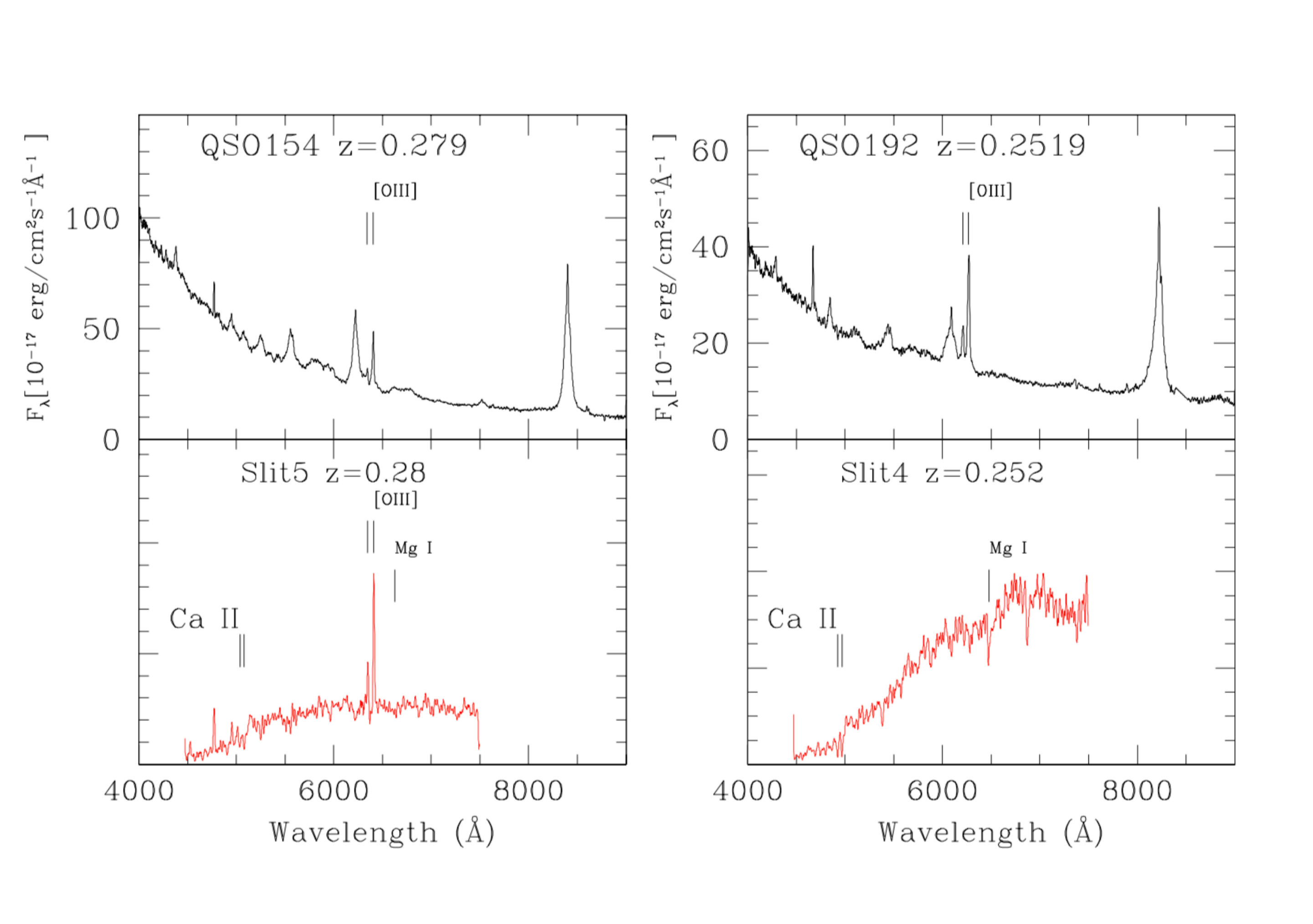}
\caption{Examples of the optical spectrum of QSO (top panels) from SDSS-DR16  and its companion galaxy (bottom panels) from this work. QSO numbers are from \protect\citet{Falomo_2014}. }
\label{fig:spectra}
\end{figure*}    

\begin{figure*}
      \includegraphics[width = 18cm]{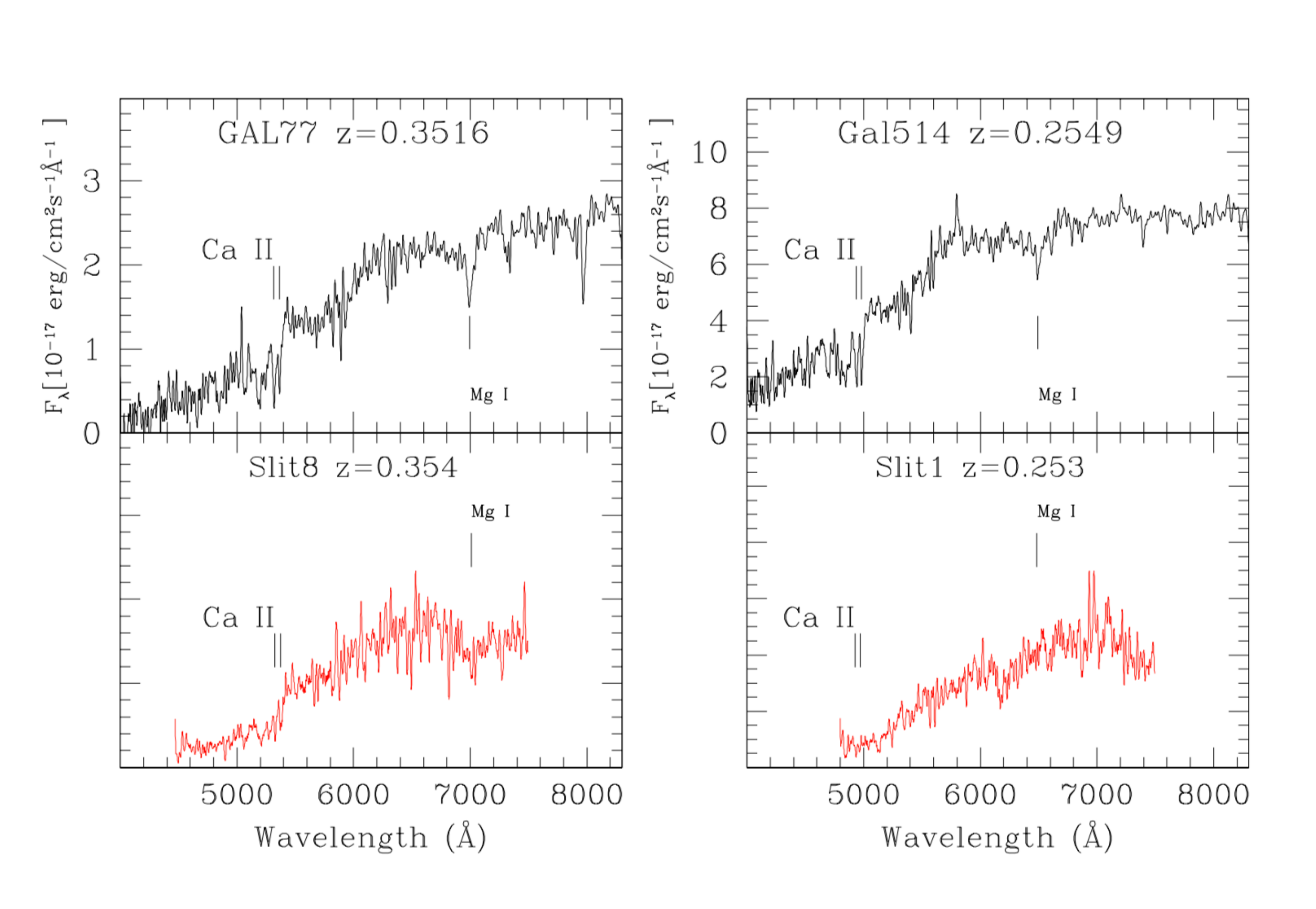}
\caption{Examples of the optical spectrum of the inactive galaxies  (top panels) from SDSS-DR16 and the companion galaxy (bottom panel) from  this work. Galaxy numbers are taken from \citet{Karhunen_2014}.}
\label{fig:spectra1}
\end{figure*}    


\begin{figure}
\centering
\includegraphics[scale=0.35]{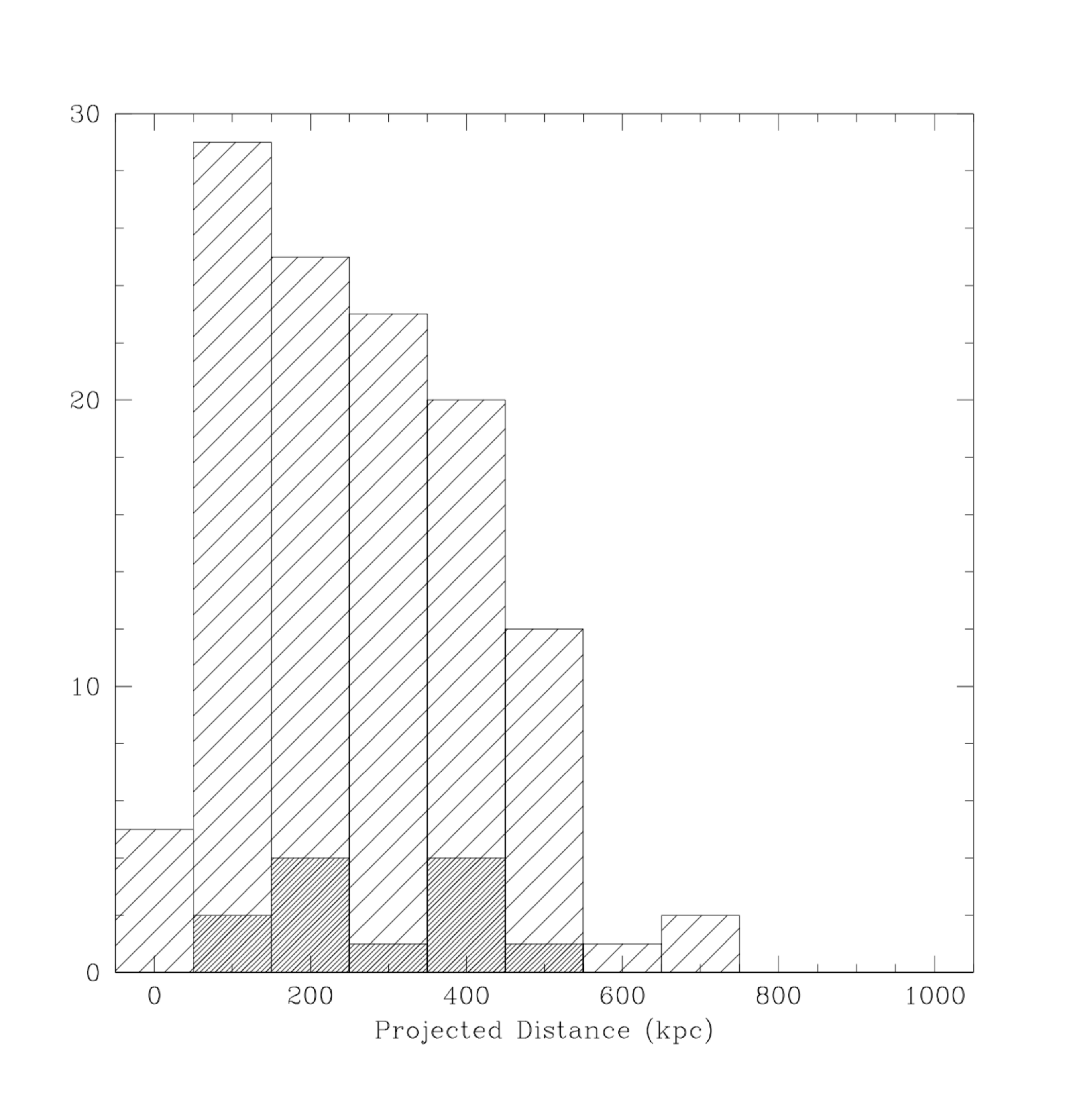}
\caption{Distribution of the projected distance (in Kpc) from the main target for observed galaxies in the fields of QSO and ING. The galaxies not associated (different redshift) to the targets are shown by shaded area while those associated (see text) are represented by filled area.}
\label{fig:comppd}
\end{figure}

\begin{figure}
\centering
\includegraphics[scale=0.35]{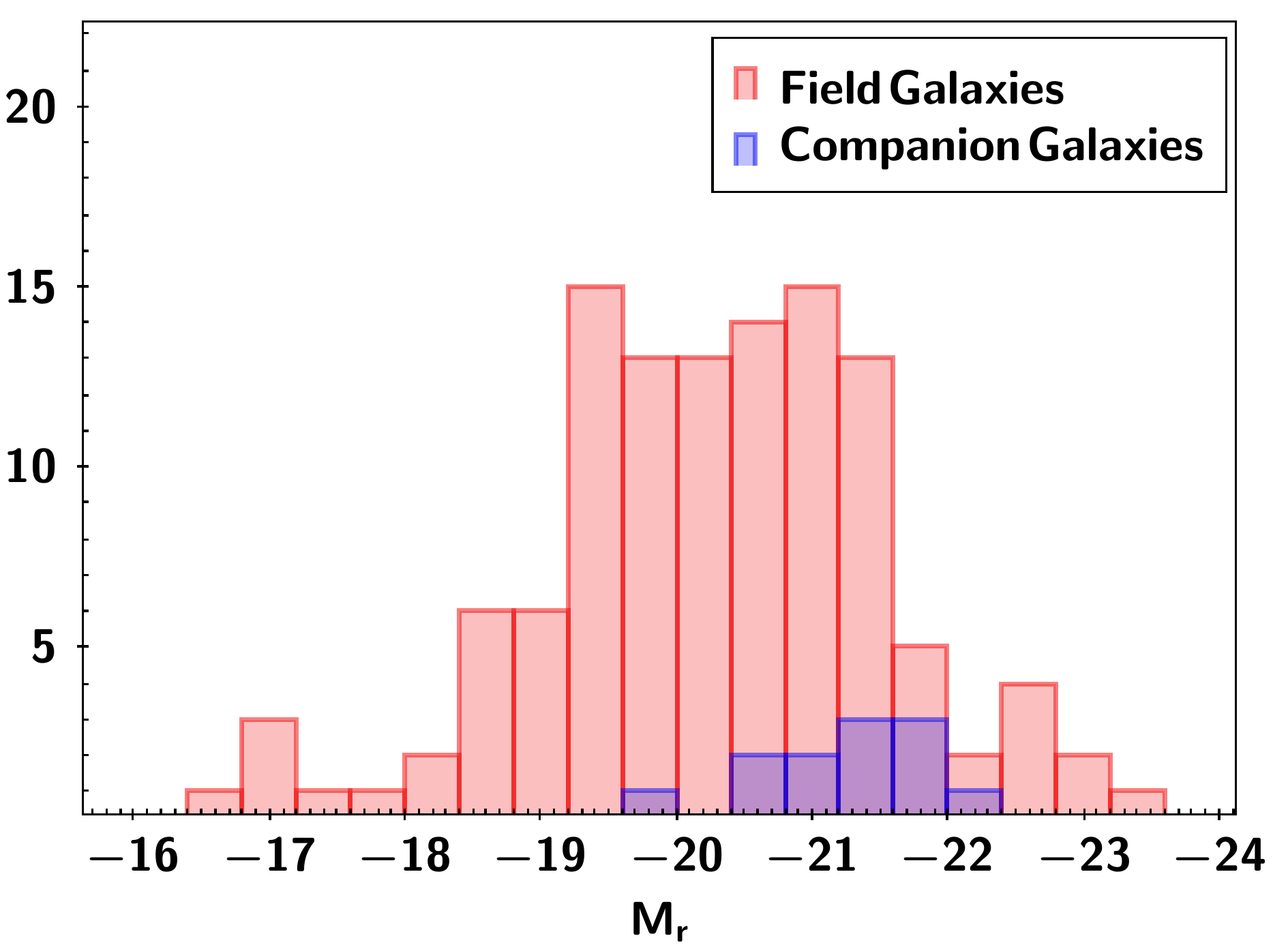}
\caption{Distribution of M$_r$ absolute magnitude of field and associated companion galaxies observed in this work.}
\label{fig:AbsoluteMagnitudes}
\end{figure}
\section{Observations and Data Analysis}

The optical spectroscopy for the galaxies in the 23 fields were secured with the ESO Faint Object Spectrograph and Camera (EFOSC2) in MOS mode at the ESO New Technology Telescope (NTT), covering the spectral range 3800-9000 \AA. We adopted the grism \#4 and a PunchHead slit width of 1.34 arcseconds, yielding an effective spectral resolution corresponding to a FWHM $\sim$ 20 \AA.
The observations were carried out in two observing runs in 2019: the first from August 28 to September 3 and the second on October 3 to 8. 
During the first observing night all the fields were observed in pre-imaging with $r$ filter with an average observing time of 20 minutes. These images were used to select the targets in each field for the subsequent MOS spectroscopic observations.
For each field therefore a slit mask of up to 10 slits of $\sim$ 8.6 arcsec length was produced.
The average brightness in $r$ band for the companion galaxies candidate in the fields is in the range $\sim$19.5-21.5. For each field we first choose the slits for the brightest targets and then, depending on the constraint on the MOS-slit positions, we selected fainter objects in order to fill all the available slits and to secure that slitlets do not overlap on the MOS plate. 
The constraints on the MOS-slit positions in some cases forced us to place the slit at the faint targets. For the 23 fields we used $\sim$ 230 slit positions. Given the above conditions, in a number of cases the spectra resulted with insufficient signal-to-noise ratio to perform reliable measurements. The same conditions were applied both for QSO fields and ING fields.  


For each MOS configuration we obtained three exposures for a total integration time of 1 hour. For 3 QSO fields (Nr 192, 205 and 239) two different MOS configurations were secured in order to optimize the selection of targets in these fields.
In Figures \ref{fig:ima} and \ref{fig:ima1} we show the isophotal contours of the fields of the QSOs and ING for which we found companions.

All the reduction steps were performed using the ESOREFLEX package \citep{Freudling_2013}. This includes flat fielding, wavelength calibration and background subtraction. 
Multiple exposures were combined with cosmic ray rejection option in order to clean the frames. 
The resulting accuracy of the wavelength calibration is 0.2 \AA. 
The 1D spectra were then extracted from the 2D background subtracted frames and calibrated for the relative sensitivity of the instrument using spectra of spectro-photometric standard stars observed during the same run. 
Finally since the use of a narrow slit might loose a fraction of the light of the companion galaxies we flux calibrated the spectra in absolute flux using the magnitude of the galaxies derived from SDSS database. 

As a first step we inspected all the final 1-D extracted spectra and keep for measurement only those with average S/N$>$5. 
The redshift was derived using the {\it rvsao} package \citep{Kurtz_1998}. The {\it emsao} routine in the cases where emission lines are present and {\it xcsao} one for pure absorption spectra. For the {\it xcsao} task we used a synthetic reference stellar spectrum of a KIII star from the \citet{Jacoby_1984} stellar library. Only spectra achieving a correlation factor $r$ \citep{Kurtz_1998} higher than 2 have been considered reliable. A further visual check was also performed. Since the correlation peak can be related to the measurement error, we estimate an average error for our measured redshift of $\sim$80 km/sec.  In 14  cases the obtained spectra turned out to be stars and were therefore discarded. At the end of this cleaning, we were able to measure a reliable redshift for 129 objects. 

\section{Results}

We are able to secure optical spectra of galaxies in the fields around 23 targets. For 12 quasar fields we measured 76 spectra and for the 11 fields around massive ING we gathered 53 spectra.
In Figures \ref{fig:spectra} and \ref{fig:spectra1} we show for two QSO and two ING the spectra of both the QSO/ING and their companion galaxy . The properties of all observed galaxies are reported in Table \ref{tab:obs_ALL}. 
 We consider that an object is associated to the QSO/ING if the difference in velocity is $\Delta V < 1000$ km/sec.
This choice represents a compromise between smaller velocity difference that ensure higher confidence of physical association but reduce the statistics and larger values that reduce the confidence but allows for better statistics. Note also that in most cases the redshift of the quasar is based (by SDSS pipeline) on all emission lines including broad lines and thus possibly biased by 500 km/s and more with respect to the systemic velocity.
The relative velocity difference cutoff limit is imposed to reasonably minimize the contamination by chance projection of companions which are not gravitation-ally bound while maintaining sufficient pair statistics \citep[][]{Patton_2000,Moreno_2013,Patton_2016,Ventou_2019}.
The measured galaxies are located at projected distance from the targets of about 20 to 700 kpc (see  Figure \ref{fig:comppd}). Those that are found associated to the targets ($\Delta V < $1000 km/s) are in the range from 50 to 500 kpc. The luminosity of the observed galaxies spreads over a wide range from M(r) =--17 to M(r)=--22 while those that are found at the same redshift of the targets cover the higher luminosity range (on average M(r)=--21.2 ) (see Figure \ref{fig:AbsoluteMagnitudes}).
The average number of galaxies detected within 0.25 Mpc of the quasar for our observed 12 QSO fields is 19$\pm$6 while for the 11 inactive galaxies fields is 18$\pm$8 \citep{Karhunen_2014}. The same similarity is found for the galaxy number density that is 7$\pm$2 ($N_{gal}/arcmin^2)$ for the 12 QSO fields and 7$\pm$3 ($N_{gal}/arcmin^2)$  for the galaxy fields \citep{Karhunen_2014}.

We found that 5 out of 12 observed fields of quasars exhibit at least one companion galaxy and in 3 cases more than one companion was found. One companion galaxy was found in 3 out of 11 ING fields. The projected separations and the luminosity of companion galaxies of quasars are similar to those of ING. Also the incidence of companions does not appear significantly different (albeit with scanty statistics) between QSO and ING in Table \ref{tab:comp} we list the data for all the companion galaxies.

As far as the star formation signature of the associated companions, we found that for QSO only one companion galaxy exhibits [OII] and/or [OIII] emission lines. On the contrary, none of the three companions of ING show emission lines\footnote{Note that one of the fields of quasars (QSO 192) was already known to have companions therefore the effective number of companions of quasars is 4 out of 11 objects.}.

\begin{table*}
    \centering
    \caption{Properties of companion galaxies from MOS observations.} 
    \label{tab:comp}
 \begin{tabular}{lcccccc}
\hline
  \multicolumn{1}{c}{id} &   \multicolumn{1}{c}{slit id} &   \multicolumn{1}{c}{SDSS} &
  \multicolumn{1}{c}{z} &   \multicolumn{1}{c}{PD} &     \multicolumn{1}{c}{$M_r$} &
  \multicolumn{1}{c}{$\Delta$V} \\
  & & & & kpc & & km/sec \\
\hline
  QSO-192 & 4 & J002834.72-000424.3 & 0.252 & 183 &  -21.37 & 50\\
   & 10 & J002825.26-000319.8 & 0.249 & 431 &   -21.03 & 869\\
  GAL-35 & 4 & J002904.22-002850.0 & 0.291 & 189 &  -20.8 & 120\\
  QSO-205 & 3 & J004324.47+005223.5 & 0.309 & 447 &  -20.53 & 210\\
   & 4 & J004322.78+005153.8 & 0.307 & 270 &  -21.6 & 390\\
  GAL-77 & 8 & J010537.57+005305.8 & 0.354 & 434 &  -22.08 & 720\\
  QSO-62 & 3 & J215744.50+005400.7 & 0.266 & 235 &  -21.72 & 420\\
  GAL-514 & 1 & J223002.81+000319.4 & 0.253 & 447 &  -21.51 & 570\\
  QSO-127 & 4 & J231253.23+001715.9 & 0.258 & 142 & -20.89 & 300\\
   & 5 & J231252.38+001754.5 & 0.26 & 169 & -19.87 & 899\\
   & 11 & J231243.31+001822.5 & 0.257 & 520 &  -21.58 & 50\\
  QSO-154 & 5 & J234933.10-003645.2 & 0.28 & 55 & -21.64 & 300\\
\hline
\end{tabular}
\end{table*}

\subsection{ Extended analysis from SDSS data}

In addition to our spectroscopic observations, we also search for companion galaxies using the SDSS DR16 \citep{Ahumada_2020} database and using the same conditions for companion selection as described above. 
 In particular, we searched the SDSS spectroscopic database for companion galaxies in the fields of the 416 QSO (\cite{Falomo_2014} and of 580 passive galaxies (\cite{Karhunen_2014}).

This search resulted in 267 QSO fields ($\sim$ 64\% of the full sample) with at least one galaxy with spectrum and 434 for the ING fields ($\sim$ 75\% of the full sample)).

\begin{figure}
\centering
\includegraphics[scale=0.35]{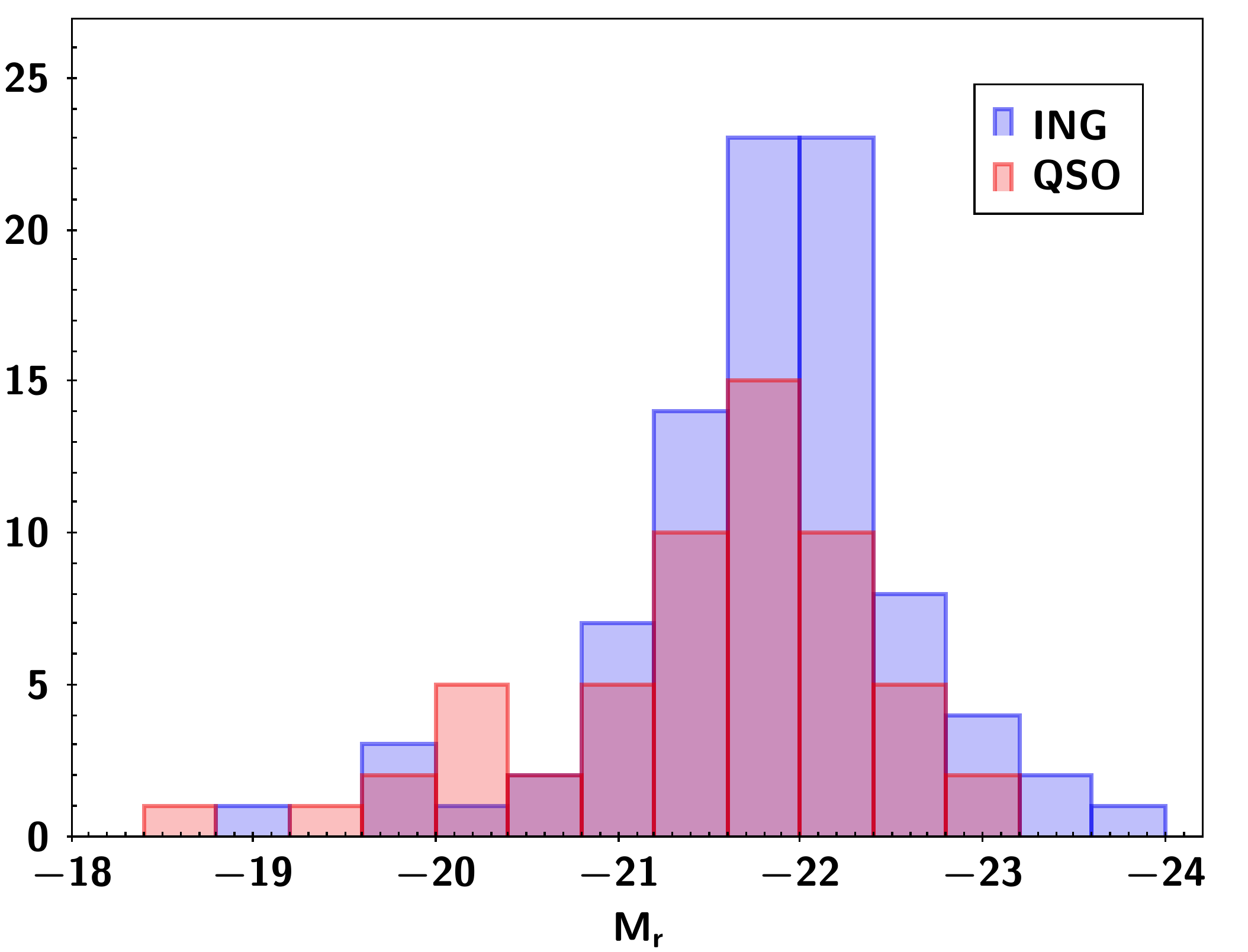}
\caption{Distribution of M$_r$ absolute magnitudes of associated companion galaxies both for QSO and ING from the extended sample derived from SDSS-DR16.}
\label{fig:comp_mag}
\end{figure}

Applying the same conditions described in Section 4 for the projected distance and the velocity difference from the target we found 58 companion galaxies in 46 QSO fields. For 34 of them at least one companion galaxy was found.  The average absolute magnitude of these companions is $<Mr >$ = -21.5 +/- 0.9. 
For the ING, we found 99 companion galaxies in 84 fields. For 62 ING at least one companion galaxy was found and for the remaining fields more than one companions was found. The average absolute magnitude of all these companions is $<Mr >$ = -21.8 +/- 0.8. The distribution of absolute magnitude is similar for the two samples (see Fig.~\ref{fig:comp_mag}). 

The incidence of companions (number of objects that have at least one companion galaxy) appears similar for the two samples:17\% for QSO and 19\% for ING. The incidence of companions of QSO from SDSS dataset is significantly lower than that found in our dataset ($\sim$ 40\%). 
Considering that the objects with SDSS spectra might be chosen somewhat differently 
from the various fields we believe that the difference of incidence of QSO companions is mainly dependent on the selection of the spectroscopic targets (long slit/MOS and fibers). In fact the distribution of the projected distance of companions is significantly different between our sample ($<PD>$ = 132 kpc) and SDSS dataset ($<PD>$ = 502 kpc). The incidence of QSO companions at PD $<$ 500 kpc for SDSS dataset is $\sim$25\% .

\section{Discussion}

Combining our new results with those from Paper \citetalias{Bettoni_2017} and Paper II, we can investigate the close environment properties of 44 low-z QSO by obtaining spectroscopy of the companion objects in their fields and probe physical association and possible events of recent star formation. For about half (19 out of 44) of the observed QSOs we found at least one associated companion galaxy and in few cases more than one associated companion is found.
In this whole sample we found 28 companion galaxies.  


 In order to better characterize the companion galaxies of QSO we evaluated the stellar mass of the 27 companions from our full sample of 44 QSO fields using the equation (1) in \cite{Gilbank_2010}, $log_{10}(M_{\star}/M_{\astrosun}) = 0.480(g-r)_0 - 0.475M_z - 0.08$ (see also \cite{Stone_2021}).
We found the average stellar mass is : $<Log(M_{\star}/M_{\astrosun}>$ = 10.6 $\pm$ 0.8.
Similarly for the 58 companions found for 46 QSO fields using the SDSS spectroscopic data (see Sect. 4.1) we found the average stellar mass is : $<Log(M_{\star}/M_{\astrosun}>$ = 11.3 $\pm$ 0.7. A moderately higher mass for the SDSS companions is expected since these companions are on average somewhat more luminous than those found in our sample ($<M(R)>$=-21.8 for SDSS companions and $<M(R)>$ = -20.9 for our sample. This is likely an observational effect because of the higher average redshift for the SDSS data ($<z>$ = 0.35 with respect to $<z>$ = 0.26 ). In fact, the spectroscopy of companion galaxies from SDSS data was secured for targets on average $\sim$ 1 magnitude brighter than the companions in our sample. This translates into more luminous objects and consequently also more massive.


Of the 19 QSO with at least one companion, 14 of them show [OII] emission lines as possible signature of recent star formation. The average [OII] luminosity for these companions is $<Log(L(OII)>=41.02\pm 0.6\times 10^{40} erg/s $. 
 
We also searched for [OII] emission line in the companion galaxies of QSO and ING from SDSS spectroscopic database (see also Section 4.1).
For the 58 companions of QSO we found spectral line measurements in SDSS dataset only for 33 objects. For the remaining 25 sources we retrieved the spectra and search for [OII] emission line. We got flux measurements of the [OII] line for 31 companion galaxies (i.e. the 53\% of the whole companion sample).
We then performed a similar search for the 99 companions of ING. Also in this case we found SDSS spectroscopic measurements only for 68 objects and for the other 31 we inspected their spectrum to search for [OII] line.
For ING there are 41 companions with [OII] line measured (i.e. 41\% of the whole companion sample).

\begin{figure}
\centering
\includegraphics[scale=0.35]{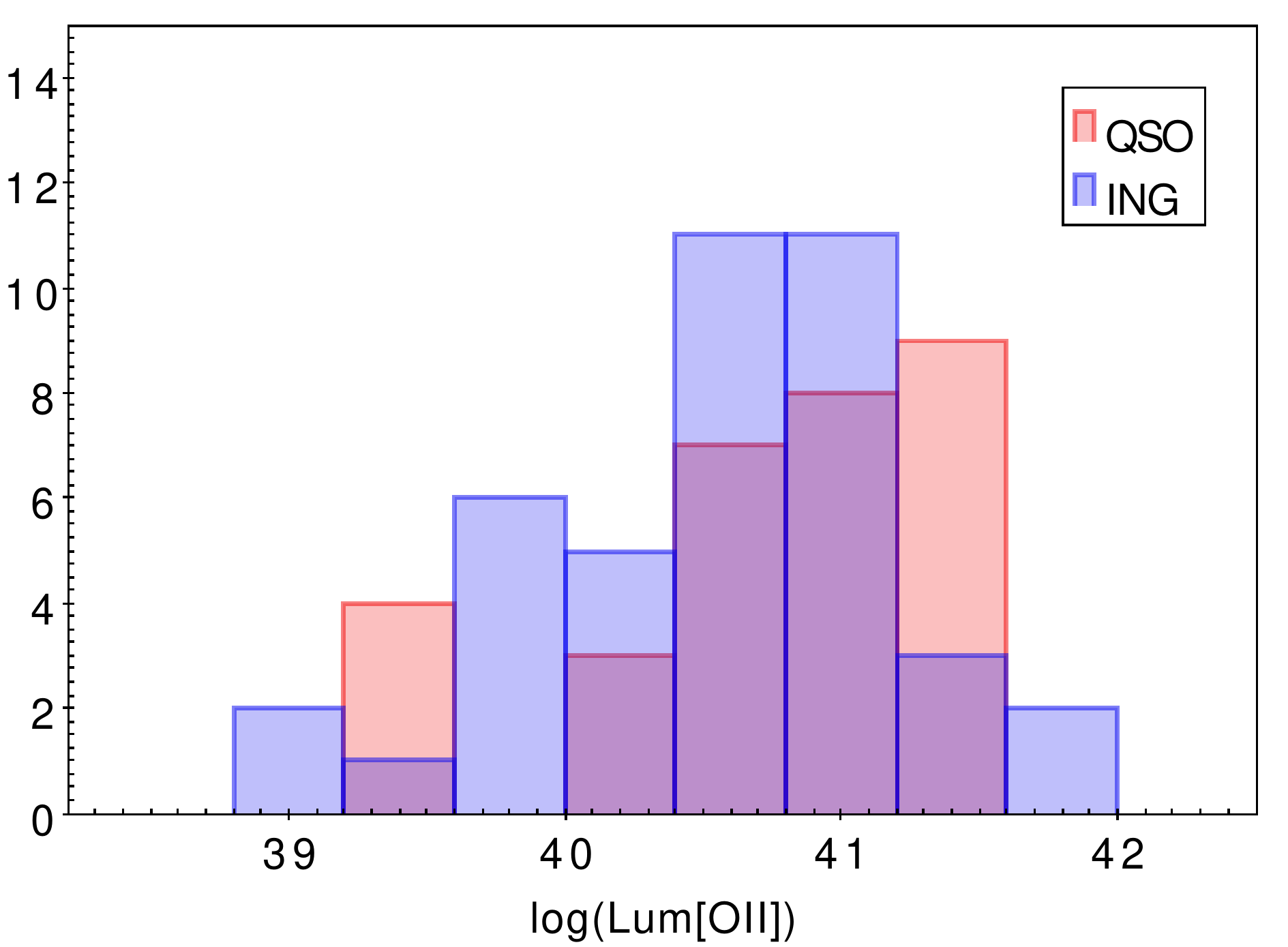}
\caption{Distribution of the [OII] 3727 \AA ~ luminosity of associated  companion galaxies both for QSO and ING derived from SDSS database. The two distributions are not significantly different (see text). }
\label{fig:comp_lumOII}
\end{figure}

Finally from these flux measurements we computed the [OII] 3727\AA \  line luminosity. We find that the distribution of the [OII] luminosity is not significantly different for companions of QSO and of ING (see Fig.~\ref{fig:comp_lumOII}). The average line luminosity is $<log(L[OII])> $ = (40.76$\pm$ 0.6 ) erg/s and $<log(L[OII])> $ = (40.49$\pm $0.6 ) erg/s for QSO and ING companions, respectively.
For both QSO and ING companions there are few objects with a particularly high star formation rate (SFR $>$ 15 $M_{\astrosun}$/year). Excluding these outliers, the two distributions are similar showing a concentration of values at SFR $<$ 4$M_{\astrosun}$/year.
On average SFR(QSO companions)= 4.9$\pm$4.0 $M_{\astrosun}$/year and SFR (ING companions) = 3.2 $\pm$2.6 $M_{\astrosun}$/year, this is comparable to the value (4.3 $M_{\astrosun}$/year)   found for the QSO companions by \citet{Stone_2021}.

This confirms our previous results \cite{Stone_2021} based on long-slit spectroscopy of 34 QSO fields.
In spite of the scanty statistics the comparison of galaxies in the fields of QSO with those of ING does not show significant differences for the incidence of close companions and/or for their global properties at low redshift. This is also consistent with the results derived from the analysis of spectroscopic data from SDSS database although in this case the incidence of companions is smaller likely due to the small number of available spectra in the fields of QSO and ING.
However, the signature of recent star formation given by the presence of emission lines of the QSO companions supports a modest link between the nuclear activity and star formation in the environments.

 Our overall results differ from the finding of \cite{Ellison_11} who report a higher incidence of close ($<$ 80 kpc) companions for a larger selection of lower redshift ($<$ 0.2)  active galactic nuclei with respect to a control sample of galaxies without companions. The higher incidence is enhanced for smaller projected separations. This suggests that the effect of companions on nuclear activity and triggering of recent star formation might be relevant only for very small physical separation of the companion from the galaxy.

 Given the significance of the topic for the investigation of the impact of the environments on the nuclear activity, it is important to extend this study to larger samples of QSO and ING with very similar characteristics of QSO host galaxies. The use of larger telescopes would also allow one to use a sample of objects at higher redshift and thus probe whether there is some evolution with the cosmic time. 
We plan to extend this program in the near future.

\section*{Acknowledgements}

We thank the anonymous referee whose comments improved the paper.
Based on observations collected at the European Southern Observatory under ESO programme(s) 0103.B-0737 and 0104.B-0197.
MBS acknowledges funding from the Finnish Cultural Foundation and Finnish Centre for Astronomy with ESO (FINCA).
MBS and JK acknowledge financial support from the Academy of Finland, grant 311438.

Funding for the Sloan Digital Sky Survey IV has been provided by the Alfred P. Sloan Foundation, the U.S. Department of Energy Office of Science and the Participating Institutions. SDSS-IV acknowledges
support and resources from the Center for High-Performance Computing at
the University of Utah. The SDSS web site is www.sdss.org.

SDSS-IV is managed by the Astrophysical Research Consortium for the 
Participating Institutions of the SDSS Collaboration including the 
Brazilian Participation Group, the Carnegie Institution for Science, 
Carnegie Mellon University, the Chilean Participation Group, the French Participation Group, Harvard-Smithsonian Center for Astrophysics, 
Instituto de Astrof\'isica de Canarias, The Johns Hopkins University, Kavli Institute for the Physics and Mathematics of the Universe (IPMU) / 
University of Tokyo, the Korean Participation Group, Lawrence Berkeley National Laboratory, 
Leibniz Institut f\"ur Astrophysik Potsdam (AIP),  
Max-Planck-Institut f\"ur Astronomie (MPIA Heidelberg), 
Max-Planck-Institut f\"ur Astrophysik (MPA Garching), 
Max-Planck-Institut f\"ur Extraterrestrische Physik (MPE), 
National Astronomical Observatories of China, New Mexico State University, 
New York University, University of Notre Dame, 
Observat\'ario Nacional / MCTI, The Ohio State University, 
Pennsylvania State University, Shanghai Astronomical Observatory, 
United Kingdom Participation Group,
Universidad Nacional Aut\'onoma de M\'exico, University of Arizona, 
University of Colorado Boulder, University of Oxford, University of Portsmouth, 
University of Utah, University of Virginia, University of Washington, University of Wisconsin, 
Vanderbilt University and Yale University.

\section*{DATA AVAILABILITY}
The data underlying this article 
are available through the ESO archive at http://archive.eso.org and can be accessed with proposal ids 0103.B-0737 and 0104.B-0197.

\bibliographystyle{mnras}
\bibliography{references.bib}    

\appendix
\section{Galaxies in the field of quasars}

\begin{table*}
    \centering
    \caption{Observed galaxies }
        \label{tab:obs_ALL}
\begin{tabular}{|l|r|l|l|c|l|r|r|r|l|}
\hline
  \multicolumn{1}{|c|}{id} &   \multicolumn{1}{c|}{slit id} &  \multicolumn{1}{c|}{SDSS} &
  \multicolumn{1}{c|}{RA} &   \multicolumn{1}{c|}{Dec} &  \multicolumn{1}{c|}{$m_r$} &
  \multicolumn{1}{c|}{PD} &   \multicolumn{1}{c|}{$M_r$} &   \multicolumn{1}{c|}{z} \\
 & & & (2000) & (2000) & & kpc &  &  \\
\hline
   GAL-32 & 3 & J002430.92+000718.1 & 6.128858 & 0.121712 &  19.25 & 349 & -22.69 & 0.44\\
   & 4 & J002430.91+000744.5 & 6.128801 & 0.129055 &  19.47 & 177 & -21.47 & 0.299\\
   & 6 & J002437.66+000840.7 & 6.156919 & 0.144652 &  18.75 & 337 & -22.08 & 0.285\\
   & 8 & J002437.22+000929.1 & 6.155118 & 0.15809 &  19.15 & 352 & -20.97 & 0.214\\
  QSO-192A & 3 & J002835.69-000308.1 & 7.14875 & -0.052264 &  21.07 & 337 & -19.34 & 0.241\\
   & 4 & J002834.72-000424.3 & 7.144705 & -0.073424 &  19.15 & 183 & -21.37 & 0.252\\
   & *6 & J002831.36-000409.8 & 7.13069 & -0.069405 &  18.75 & -- & -- & --\\
   & 7 & J002829.81-000311.3 & 7.124212 & -0.05314 &  19.05 & 241 & -21.12 & 0.218\\
   & 8 & J002828.98-000333.1 & 7.120756 & -0.059209 & 19.25 & 202 & -20.91 & 0.218\\
   & 9 & J002826.55-000258.5 & 7.110658 & -0.049595 &  21.3 & 488 & -19.72 & 0.308\\
   & 10 & J002825.26-000319.8 & 7.105283 & -0.055525 &  19.46 & 431 & -21.03 & 0.249\\
 QSO-192B  & 3 & J002835.57-000544.7 & 7.148225 & -0.095774 &  19.6 & 673 & -22.75 & 0.517\\
   & 4 & J002834.47-000452.6 & 7.143654 & -0.081303 & 20.74 & 362 & -22.41 & 0.229\\
   & 5 & J002832.86-000446.7 & 7.136934 & -0.07964 &  20.26 & 132 & -19.9 & 0.218\\
   & 7 & J002829.37-000248.6 & 7.122386 & -0.046845 &  21.56 & 326 & -18.63 & 0.22\\
   & 8 & J002828.72-000341.7 & 7.119697 & -0.061607 &  18.73 & 193 & -21.43 & 0.218\\
   & *10 & J002824.70-000453.8 & 7.10293 & -0.081622 &  17.51 & -- & -- & --\\
  GAL-35 & 4 & J002904.22-002850.0 & 7.267615 & -0.480557 &  20.08 & 189 & -20.8 & 0.291\\
   & 7 & J002900.06-002825.8 & 7.250284 & -0.47386 &  18.03 & 68 & -21.49 & 0.167\\
   & 8 & J002858.12-002730.8 & 7.242195 & -0.458572 &  20.32 & 345 & -20.49 & 0.283\\
   & 9 & J002856.44-002723.1 & 7.235168 & -0.45644 & 19.78 & 485 & -21.35 & 0.321\\
   & 10 & J002854.14-002815.1 & 7.225603 & -0.470888 & 21.38 & 454 & -19.21 & 0.259\\
  QSO-195 & 2 & J003438.06-001301.1 & 8.658602 & -0.216976 &  19.36 & 465 & -21.92 & 0.342\\
   & 4 & J003435.29-001401.0 & 8.647048 & -0.233638 &  19.0 & 281 & -21.5 & 0.249\\
   & 7 & J003431.17-001306.4 & 8.629897 & -0.218455 & 20.83 & 51 & -20.42 & 0.338\\
   & 9 & J003428.43-001326.4 & 8.618481 & -0.224004 &  19.11 & 146 & -20.39 & 0.165\\
  GAL-45 & 6 & J004045.79+000848.6 & 10.190822 & 0.146835 &  20.61 & 45 & -20.05 & 0.267\\
   & 8 & J004049.02+000833.9 & 10.204253 & 0.142758 &  21.13 & 225 & -19.28 & 0.241\\
  QSO-204 & 5 & J004215.17+005904.7 & 10.563217 & 0.984665 &  19.51 & 119 & -21.18 & 0.27\\
   & 6 & J004213.73+005804.9 & 10.55725 & 0.968047 &  20.15 & 172 & -18.83 & 0.133\\
   & 7 & J004212.89+005905.2 & 10.553716 & 0.984802 &  20.37 & 82 & -21.33 & 0.403\\
   & 8 & J004211.54+005739.3 & 10.548116 & 0.960936 &  18.44 & 515 & -22.97 & 0.359\\
  QSO-205A & 1 & J004327.59+005009.5 & 10.864977 & 0.835977 &  21.06 & 705 & -20.51 & 0.382\\
   & 3 & J004324.47+005223.5 & 10.851991 & 0.873201 &  20.5 & 448 & -20.53 & 0.309\\
   & 4 & J004322.78+005153.8 & 10.844932 & 0.864971 &  19.41 & 270 & -21.6 & 0.307\\
   & 6 & J004320.54+005059.9 & 10.8356 & 0.849998 &  20.1 & 67 & -19.95 & 0.208\\
   & 7 & J004318.07+005117.2 & 10.825316 & 0.854799 &  20.39 & 116 & -20.72 & 0.318\\
   & 9 & J004314.63+005138.3 & 10.810969 & 0.860662 &  20.47 & 190 & -18.53 & 0.134\\
 QSO-205B   & 1 & J004326.98+005243.8 & 10.862419 & 0.878852 &  22.2 & 360 & -17.01 & 0.147\\
   & 2 & J004325.80+005125.6 & 10.857539 & 0.857129 &  22.04 & 327 & -18.15 & 0.221\\
   & 3 & J004324.33+005127.8 & 10.851384 & 0.85774 &  21.63 & 133 & -16.75 & 0.103\\
   & 6 & J004319.77+005155.0 & 10.832392 & 0.865293 &  20.52 & 136 & -19.56 & 0.21\\
  GAL-55 & 2 & J005029.39-002211.5 & 12.622493 & -0.369876 &  20.31 & 324 & -20.06 & 0.237\\
   & 3 & J005027.76-002400.6 & 12.615691 & -0.400179 &  20.37 & 389 & -20.74 & 0.319\\
   & 5 & J005025.06-002318.7 & 12.604427 & -0.388548 &  19.17 & 77 & -20.39 & 0.315\\
   & 6 & J005023.83-002224.9 & 12.599326 & -0.373584 &  20.37 & 120 & -20.18 & 0.255\\
   & 7 & J005022.28-002238.6 & 12.592871 & -0.377411 &  21.39 & 116 & -18.55 & 0.199\\
    & *8 & J005019.91-002322.4 & 12.58298 & -0.389557 &  15.32 & -- & -- & -- \\
  GAL-76 & 1 & J010517.21+005203.3 & 16.321709 & 0.867587 &  19.98 & 424 & -20.49 & 0.246\\
   & 2 & J010513.92+005228.5 & 16.308042 & 0.874592 & 21.3 & 418 & -19.52 & 0.284\\
   & 3 & J010515.37+005238.8 & 16.31406 & 0.877458 &  21.09 & 298 & -19.31 & 0.24\\
   & 4 & J010515.22+005308.5 & 16.313437 & 0.885696 &  21.17 & 179 & -18.86 & 0.206\\
   & 5 & J010521.32+005319.1 & 16.338868 & 0.888654 &  20.64 & 207 & -18.95 & 0.172\\
   & *7 & J010515.78+005423.8 & 16.31575 & 0.906617 &  17.95 & -- & -- & --\\
   & 8 & J010517.29+005444.0 & 16.322064 & 0.912239 &  19.1 & 202 & -21.43 & 0.253\\
   & 9 & J010516.81+005458.5 & 16.320042 & 0.916264 &  19.93 & 197 & -19.72 & 0.176\\
  & 10 & J010513.59+005529.7 & 16.306626 & 0.92494 & 20.35 & 357 & -19.53 & 0.194\\
  GAL-77 & 1 & J010550.20+005308.2 & 16.459191 & 0.885623 &  20.5 & 488 & -20.67 & 0.327\\
   & 3 & J010546.93+005313.4 & 16.445542 & 0.887064 &  20.15 & 244 & -20.86 & 0.307\\
   & 5 & J010543.84+005300.9 & 16.43268 & 0.883587 & 19.63 & 64 & -20.73 & 0.236\\
   & 8 & J010537.57+005305.8 & 16.406576 & 0.884954 &  19.3 & 434 & -22.08 & 0.354\\
\hline\end{tabular}
\end{table*}

  \begin{table*}
    \centering
    \caption{Observed galaxies; continue}
        \label{tab:obs_ALL}
        \begin{tabular}{|l|r|l|l|c|l|r|r|r|l|}
\hline
  \multicolumn{1}{|c|}{id} &
  \multicolumn{1}{c|}{slit id} &
  \multicolumn{1}{c|}{SDSS} &
  \multicolumn{1}{c|}{RA} &
  \multicolumn{1}{c|}{Dec} &
   \multicolumn{1}{c|}{$m_r$} &
  \multicolumn{1}{c|}{PD} &
  \multicolumn{1}{c|}{$M_r$} &
  \multicolumn{1}{c|}{z} \\
   & & & (2000) & (2000) &  & kpc &  &  \\
\hline
     GAL-80 & 4 & J010611.40-001939.1 & 16.54751 & -0.327541 &  20.19 & 151 & -21.11 & 0.343\\
   & 5 & J010615.18-001912.1 & 16.563272 & -0.320034 &  19.45 & 220 & -21.34 & 0.28\\
   & 6 & J010611.51-001850.4 & 16.547963 & -0.314 &  18.76 & 54 & -20.81 & 0.17\\
   & 7 & J010611.25-001821.8 & 16.546884 & -0.30606 &  19.47 & 137 & -20.09 & 0.169\\
   GAL-81 & 3 & J010616.20+005515.1 & 16.567531 & 0.92088 &  19.39 & 287 & -21.2 & 0.258\\
      & 4 & J010608.49+005550.4 & 16.535408 & 0.930693 &  18.18 & 402 & -23.31 & 0.371\\
      & 6 & J010609.71+005641.4 & 16.540485 & 0.944856 &  19.18 & 240 & -21.25 & 0.242\\
      & 8 & J010611.27+005720.5 & 16.546962 & 0.955698 &  19.15 & 256 & -20.94 & 0.211\\
  QSO-239A & 2 & J012054.02-001942.1 & 20.22509 & -0.328372 &  21.52 & 292 & -18.63 & 0.217\\
   & 3 & J012050.51-001919.5 & 20.210488 & -0.322097 &  20.77 & 173 & -19.53 & 0.23\\
   & 3 & J012050.51-001919.5 & 20.210487 & -0.322097 &  20.77 & 376 & -20.35 & 0.323\\
   & 4 & J012049.56-001849.0 & 20.206527 & -0.313618 &  21.66 & 96 & -18.62 & 0.229\\
   & 5 & J012050.27-001829.4 & 20.209467 & -0.308185 &  19.73 & 52 & -21.56 & 0.342\\
   & 6 & J012050.84-001820.7 & 20.211846 & -0.305758 &  19.12 & 49 & -21.41 & 0.253\\
   & 8 & J012051.73-001733.4 & 20.215565 & -0.29262 &  20.28 & 157 & -18.95 & 0.148\\
   & 9 & J012051.28-001712.4 & 20.213697 & -0.286787 &  18.68 & 174 & -20.06 & 0.12\\
   & 10 & J012055.53-001653.8 & 20.231412 & -0.281628 &  19.37 & 470 & -21.11 & 0.248\\
 QSO-239B  & 1 & J012055.05-002011.6 & 20.229396 & -0.336566 &  20.42 & 384 & -19.54 & 0.2\\
   & 2 & J012053.83-001926.9 & 20.224309 & -0.324165 &  19.49 & 251 & -20.76 & 0.226\\
   & 4 & J012051.32-001844.9 & 20.213844 & -0.312487 &  20.74 & 52 & -19.79 & 0.252\\
   & 6 & J012052.51-001819.3 & 20.218821 & -0.305365 &  20.91 & 101 & -19.43 & 0.233\\
   & 7 & J012055.53-001740.6 & 20.231381 & -0.294636 &  20.48 & 337 & -20.01 & 0.249\\
   & 8 & J012045.29-001714.3 & 20.188725 & -0.287309 &  19.02 & 591 & -22.45 & 0.369\\
   & 9 & J012055.53-001653.8 & 20.231412 & -0.281628 &  19.37 & 470 & -21.12 & 0.249\\
  GAL-120 & 4 & J014651.96+001218.5 & 26.716505 & 0.205162 &  20.12 & 43 & -19.77 & 0.194\\
   & 5 & J014652.66+001231.8 & 26.719458 & 0.208836 &  20.32 & 17 & -17.32 & 0.075\\
  QSO-270 & 2 & J015243.47+000832.5 & 28.181137 & 0.142384 &  20.89 & 369 & -19.67 & 0.256\\
   & *3 & J015242.48+000750.1 & 28.177038 & 0.130607 &  19.43 & -- & -- & --\\
   & 6 & J015237.44+000832.1 & 28.156018 & 0.142255 &  18.66 & 278 & -22.03 & 0.269\\
   & 8 & J015233.45+000933.7 & 28.139416 & 0.159375 &  20.17 & 305 & -20.16 & 0.233\\
   & 1 & J015245.20+000940.9 & 28.188347 & 0.161374 &  21.15 & 456 & -20.11 & 0.339\\
   & 2 & J015243.68+000853.0 & 28.182022 & 0.148076 &  20.3 & 382 & -20.78 & 0.315\\
   & 5 & J015238.62+000858.6 & 28.160924 & 0.149639 &  21.05 & 92 & -18.05 & 0.14\\
   & *8 & J015232.47+001102.4 & 28.13532 & 0.184005 &  20.52 & -- & -- & --\\
  QSO-16 & *1 & J205002.39-001158.0 & 312.5099 & -0.19945 &  18.49 & -- & -- & --\\
   & 3 & J204959.77-001042.1 & 312.499056 & -0.178361 & 20.13 & 454 & -21.18 & 0.345\\
   & 4 & J204957.82-001135.8 & 312.490936 & -0.193282 &  19.76 & 109 & -20.34 & 0.212\\
   & *7 & J204953.28-001219.3 & 312.472 & -0.20536 &  18.97 & -- & -- & --\\
   & *9 & J204950.51-001032.4 & 312.46049 & -0.17568 &  17.15 & -- & -- & --\\
   & 10 & J204949.21-001128.3 & 312.455045 & -0.191208 &  20.34 & 364 & -19.46 & 0.188\\
  QSO-62 & 1 & J215741.54+005448.4 & 329.423108 & 0.913448 &  20.44 & 398 & -19.75 & 0.22\\
   & *2 & J215743.40+005417.9 & 329.43084 & 0.90497 &  17.56 & -- & -- & --\\
   & 3 & J215744.50+005400.7 & 329.435436 & 0.900209 &  18.94 & 235 & -21.72 & 0.266\\
   & 5 & J215744.26+005313.8 & 329.434427 & 0.887169 &  19.79 & 44 & -21.06 & 0.288\\
   & 9 & J215743.39+005139.8 & 329.430827 & 0.86108 &  19.37 & 419 & -21.98 & 0.351\\
  QSO-68 & *1 & J215955.74+000920.3 & 329.98228 & 0.15564 &  15.56 & -- & -- & --\\
   & 2 & J215954.23+000947.5 & 329.975959 & 0.163198 &  21.17 & 340 & -19.61 & 0.28\\
   & 3 & J215953.15+000842.2 & 329.971459 & 0.145063 &  19.88 & 515 & -21.5 & 0.355\\
   & 4 & J215951.87+000949.6 & 329.966125 & 0.163793 &  20.02 & 180 & -20.53 & 0.254\\
   & 8 & J215946.16+000946.5 & 329.942337 & 0.162929 &  20.81 & 357 & -22.89 & 0.86\\
  GAL-514 & 1 & J223002.81+000319.4 & 337.511713 & 0.055415 &  19.02 & 447 & -21.51 & 0.253\\
   & 3 & J222959.39+000421.3 & 337.497473 & 0.072591 &  20.67 & 188 & -19.76 & 0.242\\
   & 5 & J222956.09+000354.7 & 337.483728 & 0.06521 &  20.79 & 62 & -19.22 & 0.205\\
   & *4 & J222957.80+000431.3 & 337.490863 & 0.075382 &  19.64 & -- & -- & --\\
   & 6 & J222954.70+000452.6 & 337.477918 & 0.081287 &  20.98 & 105 & -17.95 & 0.13\\
   & *9 & J222950.41+000547.0 & 337.46005 & 0.096401 &  18.88 & -- & -- & --\\
   & 10 & J222949.54+000254.5 & 337.456454 & 0.04849 & 20.52 & 396 & -19.27 & 0.187\\
  QSO-127 & 2 & J231257.16+001643.9 & 348.238187 & 0.278865 &  19.26 & 459 & -21.78 & 0.31\\
   & 4 & J231253.23+001715.9 & 348.221801 & 0.287771 &  19.7 & 142 & -20.89 & 0.258\\
   & 5 & J231252.38+001754.5 & 348.21827 & 0.298477 &  20.74 & 169 & -19.87 & 0.26\\
   & 7 & J231249.62+001751.0 & 348.206766 & 0.297515 &  20.82 & 182 & -20.47 & 0.343\\
  & 8 & J231248.03+001839.0 & 348.200162 & 0.310856 &  19.04 & 305 & -20.97 & 0.204\\
   & 11 & J231243.31+001822.5 & 348.180481 & 0.306263 &  19.0 & 520 & -21.58 & 0.257\\
\hline\end{tabular}
\end{table*}

  \begin{table*}
    \centering
    \caption{Observed galaxies; continue}
        \label{tab:obs_ALL}
\begin{tabular}{|l|r|l|l|c|l|r|r|r|l|}
\hline
  \multicolumn{1}{|c|}{id} &
  \multicolumn{1}{c|}{slit id} &
  \multicolumn{1}{c|}{SDSS} &
  \multicolumn{1}{c|}{RA} &
  \multicolumn{1}{c|}{Dec} &
   \multicolumn{1}{c|}{$m_r$} &
  \multicolumn{1}{c|}{PD} &
  \multicolumn{1}{c|}{$M_r$} &
  \multicolumn{1}{c|}{z} \\
   & & & (2000) & (2000) &  & kpc &  &  \\
\hline
  GAL-548 & 1 & J232053.26-001943.3 & 350.221938 & -0.328719 &  21.4 & 369 & -18.93 & 0.233\\
   & 2 & J232054.49-001923.4 & 350.227065 & -0.323177 &  18.37 & 238 & -21.07 & 0.161\\
   & 3 & J232050.86-001901.8 & 350.211921 & -0.317185 &  21.1 & 183 & -18.66 & 0.184\\
   & 4 & J232051.79-001850.3 & 350.215795 & -0.31398 &  20.47 & 162 & -19.77 & 0.225\\
   & 4 & J232051.85-001844.8 & 350.216068 & -0.312452 &  20.02 & 120 & -17.1 & 0.06\\
   & 5 & J232051.20-001827.5 & 350.213358 & -0.30764 &  20.62 & 80 & -19.15 & 0.185\\
   & 6 & J232053.20-001803.3 & 350.221701 & -0.300929 &  20.82 & 62 & -19.45 & 0.228\\
   & 7 & J232051.45-001751.2 & 350.214379 & -0.297556 &  21.0 & 80 & -20.09 & 0.316\\
   & *8 & J232050.85-001731.8 & 350.21191 & -0.29217 &  17.39 & -- & -- & --\\
   & 9 & J232050.07-001652.8 & 350.208643 & -0.281351 &  21.48 & 331 & -19.27 & 0.275\\
   & 11 & J232053.54-001614.9 & 350.223102 & -0.270816 &  19.92 & 450 & -20.64 & 0.256\\
  QSO-147 & 2 & J234140.96-003930.4 & 355.420669 & -0.658461 &  18.71 & 347 & -21.97 & 0.268\\
   & 5 & J234142.76-003834.8 & 355.42818 & -0.643011 &  20.51 & 123 & -19.75 & 0.226\\
   & 6 & J234138.16-003808.2 & 355.409032 & -0.635635 &  18.94 & 206 & -21.73 & 0.268\\
   & 10 & J234143.23-003641.0 & 355.430157 & -0.611413 &  20.69 & 339 & -19.7 & 0.239\\
  QSO-154 & 3 & J234936.33-003538.1 & 357.401411 & -0.593943 &  18.73 & 276 & -21.14 & 0.192\\
   & 5 & J234933.10-003645.2 & 357.38794 & -0.612579 &  19.12 & 54 & -21.64 & 0.28\\
   & 8 & J234926.74-003614.1 & 357.361437 & -0.603933 &  19.56 & 359 & -20.79 & 0.236\\
   & 9 & J234925.54-003753.1 & 357.356445 & -0.631443 & 21.54 & 251 & -16.94 & 0.107\\
\hline\end{tabular}
     \begin{list}{}{}
     \item[]Column (1) catalog identification number from \cite{Falomo_2014}, column (3) slit identification, an asterisk indicates that the object measured turned out to be a star, column (3) SDSS identifier, column (4) object RA. column (5) object Dec, column (6) apparent $r$ magnitude, column (7) Projected Distance from the QSO/ING in kpc, column (8) absolute $r$ magnitude and column (9) measured redshift.
    \end{list} 
\end{table*}

\renewcommand{\thefigure}{\arabic{figure} (Cont.)}
\addtocounter{figure}{-1}



\bsp	
\label{lastpage}
\end{document}